# A New Approach to Decoding of Rational Irreducible Goppa code

Ahmed DRISSI
LabSiv : Laboratoire des Systèmes informatiques et Vision
ESCAM : Equipe de la Sécurité, Cryptographie, Contrôle d'Accès et Modélisation
Departments of Mathematics and Computer Science
Faculty of Sciences, Ibn Zohr University, Agadir, Morocco
idrissi2006@yahoo.fr

Ahmed ASIMI
LabSiv : Laboratoire des Systèmes informatiques et Vision
ESCAM : Equipe de la Sécurité, Cryptographie, Contrôle d'Accès et Modélisation
Departments of Mathematics and Computer Science
Faculty of Sciences, Ibn Zohr University, Agadir, Morocco
asimiahmed2008@gmail.com

*Abstract*— **The interesting properties of classical Goppa code and its effective decoding algorithm (algorithm of patterson) make the most appropriate candidate for use in the MC Eliece cryptosystem. Information leakage which results from the relationship between the error vector weight and the number of iterations in the decoding algorithm, presented a weakness of the cryptosystem. In this paper, we introduce a new approach to decoding, the use of binary Goppa code in system design MC Eliece which solve the problem of the leak of information, on the contrary in case of patterson algorithm. We treat this decoding method using the Newton identities and results of linear algebra.**

## I-INTRODUCTION

A motivation of this work is to find algorithms for decoding binary Goppa code where their use in the design of the MC Eliece leaves no information leakage. To attack the system Mc Eliece, the researchers H.Gregor Molter.Marc Stottinger.Abdulhadi Shoufan.Falko Strenzke have exploited in [1] an information leak, which results from the relationship between the weight error vector and the number of iterations of the Euclidean algorithm extended used in the algorithm of Patterson, and extract the error vector which is secret, and thereafter the plaintext. it prompts us to seek another decoding algorithm where this leak information about the error is remedied. Magali Bardet used in [2] and [3] the Newton identities for decoding cyclic codes but also used the Grobner basis calculation and the theory of elimination. The similarity in structure between the control matrix of a cyclic code and a Goppa code has encouraged us to try to follow the same path, but it has happened that the use of the properties of circular matrices and diagonalization better for our code.

In the next section, we state the notations used in this document and the third we define the Goppa code binary, its characterization and its correction capability. For the fourth section, it was replaced problem of solving a system in $F_{2^m}^n$ to $m$ systems in $F_2^n$ . And its resolutions are discussed in the following two sections treating the relationship between Newton and elementary symmetric functions. Transforming this relationship in matrix form and use the properties of linear algebra, in particular the structure of a circulant matrix. We finally give our own method for decoding a binary irreducible Goppa code.

## II- NOTATIONS AND PRELIMINARIES

$m$ : An integer.

$F_{2^m}$ : A finite field of $2^m$ elements.

$F_2 = \{0,1\}$.

$F_2^n$ : The set of vectors of length $n$ of a component 0 or 1.

$F_{2^m}^n$ : The set of vectors of length $n$ a component of $F_{2^m}$ .

$I_n$ : The identity matrix of size $n$ .

$I$ : an identity matrix.

$I_d$ : The application identity.

$mat_\beta(f)$ : The matrix associated with the endomorphism $f$ in the base $\beta$ .

$\frac{d\sigma_a(x)}{dx}$ : The derivative of the polynomial $\sigma_a(x)$ relative to $x$.

$\Gamma(L,g)$ : The Goppa code of support $L$ and polynomial $g$.

$[\ ]$ : The integer part.

$(w_1,...,w_m)$ : the basis of the vector space $F_{2^m}$ on the field $F_2$.

$F_{2^m}[x]$ : The set of polynomials with coefficients in $F_{2^m}$.

$N$ : The set of integers.

$card$ : the number of elements of a set.

$rg(C)$ : the rank of a matrix $C$

Let $\alpha \in F_{2^m}$ and

$g(x) = g_0 + g_1 x + ... + g_r x^r \in F_{2^m}[x]$ with

$g_r \neq 0$ and $g(\alpha) \neq 0$;

$$g(x) - g(\alpha) = g_1(x-\alpha) + ... + g_r(x^r - \alpha^r) = (x-\alpha)\sum_{k=1}^{r} g_k \sum_{j=0}^{k-1} x^j \alpha^{k-1-j}$$

.

Therefore

$$\frac{-g(x)}{g(\alpha)} + 1 = (x-\alpha).\left[-\frac{1}{g(\alpha)}\sum_{k=1}^{r} g_k\left(\sum_{i=0}^{k-1} x^i \alpha^{k-1-i}\right)\right];$$

It is said that

$$\frac{1}{x-\alpha} \bmod g(x) = -\frac{1}{g(\alpha)}\sum_{k=1}^{r} g_k\left(\sum_{i=0}^{k-1} x^i \alpha^{k-1-i}\right).$$

## III- The Binary Goppa code

### *1-Definition*

Let $L = (\alpha_1,...,\alpha_n)$ a sequence of $n$ distinct elements of $F_{2^m}$ and $g(x) \in F_{2^m}[x]$ a polynomial of degree $r$ in $F_{2^m}[x]$ such as $1 \le r \le n-1$ and $g(\alpha_i) \neq 0$ for all $i = 1,.....,n$.

Rational Goppa code of support $L$ (vector generator) and of generator polynomial $g$ ( Goppa polynomial) noted $\Gamma(L,g)$ is the set

$$\Gamma(L,g) = \left\{ c = (c_1,...,c_n) \in F_2^n / \sum_{i=1}^{n} c_i\left(\frac{1}{x-\alpha_i} \bmod g(x)\right) = 0 \right\}$$

If $g$ is irreducible, we say that $\Gamma(L,g)$ is an irreducible binary Goppa code.

### *2- Characterization of Goppa code*

***Theorem***

Let $L = (\alpha_1,...,\alpha_n)$ a sequence of $n$ distinct elements of $F_{2^m}$ and $g(x) \in F_{2^m}[x]$ a polynomial of degree $r$ in $F_{2^m}[x]$, such as $1 \le r \le n-1$ and $g(\alpha_i) \neq 0$ for all $i = 1,.....,n$.

The following assertions are equivalent:

$i)$ $\sum_{i=1}^{n} a_i\left(\frac{1}{x-\alpha_i} \bmod g(x)\right) = 0$.

$ii)$ $Ha^t = 0$ with

$$H = \begin{pmatrix} 1 & 1 & ... & 1 \\ \alpha_1 & \alpha_2 & ... & \alpha_n \\ .. & .... & ... & ... \\ \alpha_1^{r-1} & \alpha_2^{r-1} & ... & \alpha_n^{r-1} \end{pmatrix} \begin{pmatrix} g(\alpha_1)^{-1} & & \\ & .. & \\ & & .. \\ & & & g(\alpha_n)^{-1} \end{pmatrix}$$

$H$ is called control matrix Goppa code $\Gamma(L,g)$

$iii)$ $g(x)$ divide $\frac{d\sigma_a(x)}{dx}$ with

$$\sigma_a(x) = \prod_{i=1}^{n} (x-\alpha_i)^{a_i}.$$

***proof***

we have

$$\frac{1}{x-\alpha} \bmod g(x) = -\frac{1}{g(\alpha)}\sum_{k=1}^{r} g_k\left(\sum_{i=0}^{k-1} x^i \alpha^{k-1-i}\right);$$

$$\begin{aligned}\sum_{i=1}^{n} a_i \frac{1}{x-\alpha_i} \bmod g(x) &= -\sum_{i=1}^{n} a_i g(\alpha_i)^{-1}\left(\sum_{k=1}^{r} g_k \sum_{j=0}^{k-1} x^{k-1-j}\alpha_i^j\right) \\ &= -\sum_{k=1}^{r} g_k \sum_{j=0}^{k-1} x^{k-1-j}\left(\sum_{i=1}^{n} a_i g(\alpha_i)^{-1}\alpha_i^j\right) \\ &= -\sum_{k=1}^{r} g_k \sum x^{k-1-j} A_j\end{aligned}$$

we denote

$$A_j = \sum_{i=1}^{n} a_i\, g(\alpha_i)^{-1}\, \alpha_i^j,\ j = 0,1,...r-1.$$

$ii) \Rightarrow i)$

$Ha^t = 0 \Rightarrow \sum_{i=1}^{n} a_i\, g(\alpha_i)^{-1}\, \alpha_i^j = 0$ for $j = 0,1,...,r-1$ then

$$\sum_{i=1}^{n} a_i \frac{1}{x-\alpha_i} \bmod g(x) = 0.$$

$i) \Rightarrow ii)$

Let $\sum_{i=1}^{n} a_i \frac{1}{x-\alpha_i} \bmod g(x) = 0$, therefore $\sum_{k=1}^{r} g_k \sum_{j=0}^{k-1} x^{k-1-j} A_j = 0$ then

$$g_1A_0 + g_2(xA_0 + A_1) + g_3(x^2A_0 + xA_1 + A_2) + g_4(x^3A_0 + x^2A_1 + xA_2 + A_3) + g_5(x^4A_0 + x^3A_1 + x^2A_2 + xA_3 + A_4) + ... + g_r(x^{r-1}A_0 + x^{r-2}A_1 + ... + xA_{r-2} + A_{r-1}) = 0$$

Then

$$g_1A_0 + g_2A_1 + g_3A_2 + ... + g_rA_{r-1} = 0$$
$$g_2A_0 + g_3A_1 + ... + g_rA_{r-2} = 0$$
$$g_3A_0 + ... + g_rA_{r-3} = 0$$
$$\dots$$
$$g_{r-2}A_0 + g_rA_1 = 0$$
$$g_rA_0 = 0.$$

Since $g_r \neq 0$, it was $A_0 = 0$.

By recurrence we find that $A_j = 0$ for $j = 0,1,....,r-1$.

$i) \Leftrightarrow iii)$

$$\frac{d\sigma_a(x)}{dx} = \sum_{i=1}^{n} a_i(x-\alpha_i)^{a_i-1}\prod_{\substack{j=1\\j\neq i}}^{n}(x-\alpha_j)^{a_j} = \sum_{i=1}^{n}\frac{a_i}{x-\alpha_i}\prod_{j=1}^{n}(x-\alpha_i)^{a_j}$$
$$= \sigma_a(x).\sum_{i=1}^{n}\frac{a_i}{x-\alpha_i} = \sigma_a(x).R_a(x).$$

It was $R_a(x) = \sum_{i=1}^{n}\frac{a_i}{x-\alpha_i} = \frac{P(x)}{Q(x)}$ with $Q(x) = \prod_{i=1}^{n}(x-\alpha_i)$ and

$$u_i(x) = \frac{1}{x-\alpha_i}\bmod g(x).$$

It was $u_i(x).(x-\alpha_i) = 1 + k_i(x).g(x)$

$$\sum_{i=1}^{n} a_i\left(\frac{1}{x-\alpha_i}\bmod g(x)\right) = \sum_{i=1}^{n} a_iu_i(x)$$
$$= \sum_{i=1}^{n}\frac{a_i}{x-\alpha_i} + \sum_{i=1}^{n}\frac{a_ik_i(x)g(x)}{x-\alpha_i}$$
$$= R_a(x) + g(x)\sum_{i=1}^{n} a_i\frac{k_i(x)}{x-\alpha_i}$$
$$= \frac{P(x)}{Q(x)} + g(x)\sum_{i=1}^{n} a_i\frac{k_i(x)}{x-\alpha_i}$$

Where

$$Q(x).\sum_{i=1}^{n} a_i\left(\frac{1}{x-\alpha_i}\bmod g(x)\right) = P(x) + g(x).Q(x).\sum_{i=1}^{n} a_i\frac{k_i(x)}{x-\alpha_i} \quad (1)$$

$\Rightarrow)$

If $\sum_{i=1}^{n} a_i\left(\frac{1}{x-\alpha_i}\bmod g(x)\right) = 0$ then $g(x)$ divide $P(x)$ (indeed $Q(x).\sum_{i=1}^{n} a_i\frac{k_i(x)}{x-\alpha_i}$ is a polynomial) gold it was $\frac{d\sigma_a(x)}{dx} = \sigma_a(x).R_a(x)$ then $Q(x).\frac{d\sigma_a(x)}{dx} = \sigma_a(x).P(x)$ therefore $g(x)$ divide $Q(x).\frac{d\sigma_a(x)}{dx}$ ; gold $Q(x)$ and $g(x)$ are mutually prime (because $g(\alpha_i) \neq 0$ for $i = 1,...,n$ ) therefore $g(x)$ divide $\frac{d\sigma_a(x)}{dx}$

$\Leftarrow)$

If $g(x)$ divide $\frac{d\sigma_a(x)}{dx}$ therefore $g(x)$ divide $Q(x).\frac{d\sigma_a(x)}{dx} = \sigma_a(x).P(x)$ and since $g(x)$ and $\sigma_a(x)$ are mutually prime therefore $g(x)$ divide $P(x)$ and according to (1) $g(x)$ divided $Q(x).\sum_{i=1}^{n} a_i\left(\frac{1}{x-\alpha_i}\bmod g(x)\right)$ and since $g(x)$ and $Q(x)$ are relatiely prime then $g(x)$ divide $\sum_{i=1}^{n} a_i\left(\frac{1}{x-\alpha_i}\bmod g(x)\right)$. We know that $\deg g(x) = r$ and

$\deg(\sum_{i=1}^{n} a_i\left(\frac{1}{x-\alpha_i}\bmod g(x)\right)) = r-1$ then

$$\sum_{i=1}^{n} a_i\left(\frac{1}{x-\alpha_i}\bmod g(x)\right) = 0$$

***3- correction capacity***

The parity matrix $H$ can be written as the product of a Vandermonde matrix and a nonsingular matrix so any square submatrix $r \times r$ of $H$ is invertible, then there is no code word of weight less or equal to $r$, then it has a minimum distance of at least $d = r + 1$ (of capacity correction $[\frac{r}{2}]$).

If we add an additional constraint on $g$ to be without multiple factors, we can double the capacity of correction. (in particular irreducible codes).

Indeed $g(x)$ divide $\frac{d\sigma_a(x)}{dx}$, gold on $F_{2^m}$ the derivative of a polynomial does not contain coefficients of odd degree, therefore there exists a polynomial satisfying $\frac{d\sigma_a(x)}{dx} = f^2(x)$.

If $g$ has no multiple factors and $g$ divide $f^2$ then, $g$ necessarily divided $f$.

A word $a$ which belongs to the code $\Gamma(L,g)$ with $g$ without multiple factors belong to the code $\Gamma(L,g^2)$ which has a minimum distance $2r+1$ and a decoding algorithm to $r$ errors.

***4- The decoding***

Formally the decoding problem can be stated as follows: let the received word $r = (r_1, r_2, ..., r_n)$ and the codeword sent $c = (c_1, c_2, ..., c_n)$ such as $r = c + e$ with $e = (e_1, e_2, ..., e_n)$ a weight vector less or equal to the correction capacity

$$\begin{pmatrix} S_0 \\ S_1 \\ \cdots \\ S_{r-1} \end{pmatrix} = \begin{pmatrix} 1 & 1 & \cdots & 1 \\ \alpha_1 & \alpha_2 & \cdots & \alpha_n \\ \cdot\cdot & \cdots & \cdots & \cdots \\ \alpha_1^{r-1} & \alpha_2^{r-1} & \cdots & \alpha_n^{r-1} \end{pmatrix} \begin{pmatrix} g(\alpha_1)^{-1} & & & \\ & \cdot & & \\ & & \cdot & \\ & & & g(\alpha_n)^{-1} \end{pmatrix} \begin{pmatrix} r_1 \\ r_2 \\ \cdots \\ r_n \end{pmatrix}$$

$$= Hr^t = Hc^t + He^t = 0 + He^t = He^t$$

$$= \begin{pmatrix} 1 & 1 & \cdots & 1 \\ \alpha_1 & \alpha_2 & \cdots & \alpha_n \\ \cdot\cdot & \cdots & \cdots & \cdots \\ \alpha_1^{r-1} & \alpha_2^{r-1} & \cdots & \alpha_n^{r-1} \end{pmatrix} \begin{pmatrix} g(\alpha_1)^{-1} & & & \\ & \cdot & & \\ & & \cdot & \\ & & & g(\alpha_n)^{-1} \end{pmatrix} \begin{pmatrix} e_1 \\ e_2 \\ \cdots \\ e_n \end{pmatrix}$$

$$= \begin{pmatrix} 1 & 1 & \cdots & 1 \\ \alpha_1 & \alpha_2 & \cdots & \alpha_n \\ \cdot\cdot & \cdots & \cdots & \cdots \\ \alpha_1^{r-1} & \alpha_2^{r-1} & \cdots & \alpha_n^{r-1} \end{pmatrix} \begin{pmatrix} g(\alpha_1)^{-1} e_1 \\ g(\alpha_2)^{-1} e_2 \\ \cdots \\ g(\alpha_n)^{-1} e_n \end{pmatrix}$$

We must find the vector $\begin{pmatrix} g(\alpha_1)^{-1} e_1 \\ g(\alpha_2)^{-1} e_2 \\ \cdots \\ g(\alpha_n)^{-1} e_n \end{pmatrix}$

$\begin{pmatrix} S_0 \\ S_1 \\ \cdots \\ S_{r-1} \end{pmatrix}$ is called the syndrome vector.

we introduce the sequence of syndromes extended $(S_i)_{i \in N}$

We see that $S_{i+2^m-1} = S_i \ \forall i \in N$ therefore we restrict to the finite sequence $S_j = \sum_{i=1}^{n} \frac{e_i}{g(\alpha_i)} \alpha_i^j$ for $j = 0,1,...,2^m - 2$.

**IV -The idendity newton**

Let $k \in N$ and $x_1, x_2, ..., x_k \in F_{2^m}$, the following theorem known as Newton's identity gives a relation between the elementary symmetric functions $\sigma_j = \sum_{1 \le i_1 < i_2 < ... < i_j \le k} x_{i_1} ... x_{i_j}$ and the sums of newton $S_p = \sum_{i=0}^{k} x_i^p$, $\forall p \in N$.

***Theorem -circular identity of Newton-***

Let $k \in N$ and $x_1, x_2, ..., x_k \in F_{2^m}$, the sum of Newton $S_i = \sum_{j=0}^{k} x_j^i$, $\forall i \in N$, the elementary symmetric functions $\sigma_1, ..., \sigma_k$ of $x_1, x_2, ..., x_k$ defined by $\sigma_j = \sum_{1 \le i_1 < i_2 < ... < i_j \le k} x_{i_1} ... x_{i_j}$ then we will have relations

$$S_i + \sum_{j=1}^{k} \sigma_j S_{i-j} = 0 \ \text{ for } i \ge k.$$

***Proof***

We denote the polynomial

$$\sigma(x) = \prod_{j=1}^{k} (x - x_j) = \sum_{j=0}^{k} \sigma_j x^{k-j} = \sigma_0 + \sigma_1 x^{k-1} + ... + \sigma_{k-1} x + \sigma_k$$

with $\sigma_0 = 1$. For $j$ fixed $\sigma(x_j) = 0$ therefore $\sigma_0 x_j^k + \sigma_1 x_j^{k-1} + ... + \sigma_{k-1} x_j + \sigma_k = 0$

Let for $p \ge k$ $x_j^{p-k} \sigma(x_j) = 0$ then $\sigma_0 x_j^p + \sigma_1 x_j^{p-1} + ... + \sigma_{k-1} x_j^{p-k+1} + \sigma_k x_j^{p-k} = 0$

Summing over $j$ we will

$$\sigma_0 S_p + \sigma_1 S_{p-1} + \ldots + \sigma_{k-1} S_{p-k+1} + \sigma_k S_{p-k} = 0$$

Since $S_{i+2^m-1} = S_i$ ( indeed $S_{i+2^m-1} = \sum_{j=1}^{k} x_j^{i+2^m-1} = \sum_{j=1}^{k} x_j^i = S_i$ )

therefore for

$1 \le p \le k$ we will $S_{p+2^m-1} = S_p$ and $p + 2^m - 1 \ge k$

then we can write this relation in matrix form as follows

***Lemma: form matrix identity newton***

Let $x_1, x_2, \ldots, x_k \in F_{2^m}$ and

$$\sigma(x) = \prod_{j=1}^{k}(x - x_j) = \sum_{j=0}^{k} \sigma_j x^{k-j} = \sigma_0 + \sigma_1 x^{k-1} + \ldots + \sigma_{k-1} x + \sigma_k$$

and

$S_i = \sum_{j=0}^{k} x_j^i$ , $\forall i = 0,1,\ldots,2^m - 2$ it was

$$\begin{bmatrix} S_{2^m-2} & S_{2^m-3} & \ldots & S_0 \\ S_0 & S_{2^m-2} & \ldots & S_1 \\ \ldots & \ldots & \ldots & \ldots \\ S_{2^m-3} & S_{2^m-4} & \ldots & S_{2^m-2} \end{bmatrix} \begin{bmatrix} 0 \\ \ldots \\ 0 \\ 1 \\ \sigma_1 \\ \ldots \\ \sigma_k \end{bmatrix} = 0 .$$

## V- The circulant matrix

***Definitions***

A circulant matrix with coefficients in a finite field $F_{2^m}$ of size $n$ is a matrix of the form $C = \begin{bmatrix} c_0 & c_1 & \ldots & c_{n-1} \\ c_{n-1} & c_0 & \ldots & c_{n-2} \\ \ldots & \ldots & \ldots & \ldots \\ c_1 & c_2 & \ldots & c_0 \end{bmatrix}$ with

$c_i \in F_{2^m} \quad \forall i \in \{0,1,\ldots,n-1\}$

The matrix $A = \begin{bmatrix} 0 & 1 & 0\ldots & 0 \\ 0 & 0 & 1 & \ldots \\ & & & ..0 \\ 0 & & \ldots 0 & 1 \\ 1 & 0 & \ldots. & 0 \end{bmatrix}$ is said to be elementary circulant matrix.

***Lemma1***

Let $A$ the circulant matrix elementary of size $n$ and $E_j$ the $j^{th}$ line of the identity matrix $I_n$ it was

$i)$ $A^n = I_n$

$ii)$ $E_j A^k = E_{(j+k) \bmod n} \quad \forall k = 1,\ldots,n$

***Proof***

$i)$ let $\beta(e_1, e_2, \ldots, e_n)$ a base and $f$ the endomorphism such as $A = mat_\beta(f)$ , we see that $f(e_1) = e_n$ and $\forall k = 2,\ldots,n$ , $f(e_k) = e_{k+1}$ , we deduce easily that $f^n = I_d$ that is to say $A^n = I_n$.

$ii)$ by recurrence on $k$ it was for $k = 1$, $E_j A = E_{j+1}$.

Suppose that $E_j A^k = E_{(j+k) \bmod n}$ then

$$E_j A^{k+1} = E_j A^k A = E_{(j+k) \bmod n} A = E_{(j+k+1) \bmod n}$$

***Lemma2***

We can decompose the circulant matrix $C$ defined above in the following manner $C = c_0 I + c_1 A + \ldots + c_{n-1} A^{n-1}$

***Proof***

Using the fact that $E_j A^k = E_{(j+k) \bmod n}$ it was

$$E_j(c_0 I + c_1 A + \ldots + c_{n-1} A^{n-1}) = c_0 E_j + c_1 E_{j+1} + \ldots + c_{n-1} E_{j+n-1} = E_j.C$$

Therefore a $j^{éme}$ line of $C$ and of $c_0 I + c_1 A + \ldots + c_{n-1} A^{n-1}$ are equal.

***Lemma3***

Let $\alpha$ be a primitive element of the finite field $F_{2^m}$ therefore $\alpha^{2^m-1} = 1$ and the matrix

$$P = \begin{bmatrix} 1 & 1 & \ldots & 1 \\ 1 & \alpha & \ldots & \alpha^{2^m-2} \\ \ldots & \ldots & \ldots & \ldots \\ 1 & \alpha^{2^m-2} & \ldots & \alpha^{(2^m-2)(2^m-2)} \end{bmatrix} = \left(\alpha^{ij}\right)_{\substack{i=0,1,..2^m-2 \\ j=0,1,\ldots,2^m-2}}$$ is

invertible and its inverse is

$$P^{-1} = \begin{bmatrix} 1 & 1 & \ldots & 1 \\ 1 & \alpha^{2^m-2} & \ldots & \alpha \\ \ldots & \ldots & \ldots & \ldots \\ 1 & \alpha^{(2^m-2)(2^m-2)} & \ldots & \alpha^{(2^m-2)} \end{bmatrix} = \left(\alpha^{-ij}\right)_{\substack{i=0,1,..2^m-2 \\ j=0,1,\ldots2^m-2}}$$

***Proof***

Let $a_{il} = \sum_{j=0}^{2^m-2} \alpha^{ij}\alpha^{-jl} = \sum_{j=0}^{2^m-2} \alpha^{j(i-l)} = \begin{cases} 1 & si \quad i = l \\ 0 & si \quad i \ne l \end{cases}$

gold it was $1 + \alpha + \alpha^2 + \ldots + \alpha^{2^m-2} = 0$ indeed

$$(1-\alpha)\left(1 + \alpha + \alpha^2 + \ldots + \alpha^{2^m-2}\right) = 1 - \alpha^{2^m-1} = 1 - 1 = 0$$

***Lemma 4***

Let $\left(c_1, c_2, ..., c_{n-1}\right) \in F_{2^m}^n$ and the polynomial

$C(x) = c_0 + c_1 x + ... + c_{n-1} x^{n-1}$ it was if $n = 2^m - 1$ and $\alpha$ primitive root of the finite field $F_{2^m}$

$$\begin{bmatrix} c_0 & c_1 & ... & c_{n-1} \\ c_{n-1} & c_0 & ... & c_{n-2} \\ ... & ... & ... & ... \\ c_1 & c_2 & ... & c_0 \end{bmatrix} = \begin{bmatrix} 1 & 1 & ... & 1 \\ 1 & \alpha & ... & \alpha^{2^m-2} \\ ... & ... & ... & ... \\ 1 & \alpha^{2^m-2} & ... & \alpha^{(2^m-2)(2^m-2)} \end{bmatrix} \begin{bmatrix} C(1) & & & \\ & C(\alpha) & & \\ & & ... & \\ & & & C(\alpha^{2^m-2}) \end{bmatrix} \begin{bmatrix} 1 & 1 & ... & 1 \\ 1 & \alpha^{2^m-2} & ... & \alpha \\ ... & ... & ... & ... \\ 1 & \alpha^{(2^m-2)(2^m-2)} & ... & \alpha^{(2^m-2)} \end{bmatrix}$$

And $rg(C) = card\left\{i \in \{0,1,...,n-1\}, C(\alpha^i) \neq 0\right\}$

***Proof***

Calculate the eigenvalues of the elementary circulate matrix $A$ of size $n$.

It was $A^{2^m-1} = I$ by Lemma 1.

Let $\alpha$ primitive root of the finite field $F_{2^m}$ (that is to say $\alpha^{2^m-1} = 1$) we have

$$\begin{bmatrix} 0 & 1 & 0... & 0 \\ 0 & 0 & 1 & ... \\ & & & ..0 \\ 0 & & ...0 & 1 \\ 1 & 0 & .... & 0 \end{bmatrix} \begin{bmatrix} 1 \\ \alpha^i \\ ... \\ \alpha^{i(2^m-2)} \end{bmatrix} = \alpha^i \begin{bmatrix} 1 \\ \alpha^i \\ ... \\ \alpha^{i(2^m-2)} \end{bmatrix} \text{ for}$$

$i = 0,1,...2^m - 2$ so we can diagonalize $A$ as follows

$$A = P \begin{bmatrix} 1 & & & \\ & \alpha & & \\ & & ... & \\ & & & \alpha^{2^m-2} \end{bmatrix} P^{-1} \text{ with}$$

$$P = \begin{bmatrix} 1 & 1 & ... & 1 \\ 1 & \alpha & ... & \alpha^{2^m-2} \\ ... & ... & ... & ... \\ 1 & \alpha^{2^m-2} & ... & \alpha^{(2^m-2)(2^m-2)} \end{bmatrix}$$

; gold it was $C = C(A)$ by *lemma 2*

$$C = C(A) = C\left( P \begin{bmatrix} 1 & & & \\ & \alpha & & \\ & & ... & \\ & & & \alpha^{2^m-2} \end{bmatrix} P^{-1} \right) = P \begin{bmatrix} C(1) & & & \\ & C(\alpha) & & \\ & & ... & \\ & & & C(\alpha^{2^m-2}) \end{bmatrix} P^{-1}$$

and we deduce that also

$$rgC = rg \begin{bmatrix} C(1) & & & \\ & C(\alpha) & & \\ & & ... & \\ & & & C(\alpha^{2^m-2}) \end{bmatrix} = card\left\{i \in \{0,1,...,n-1\}, C(\alpha^i) \neq 0\right\}.$$

## VI- FOR $F_{2^m}$ TO $F_2^m$

we put $v = \left[v_1, ..., v_n\right] \in F_{2^m}^n$ and we must solve the following system in $F_{2^m}^n$ : $Av^t = S$ with the matrix $A$ and the vector $S$ are known.

In this section we will replace this system by $m$ systems unknown in $F_2^n$

$$Av^\lambda = S^\lambda, \lambda = 1, ..., m$$

Let $\left(\omega_1, ..., \omega_m\right)$ a base of $F_{2^m}$ as a vector space on field $F_2$

$$\forall v \in F_{2^m} ; v = \sum_{\lambda=1}^{m} v_\lambda \omega_\lambda \text{ and } \forall s \in F_{2^m} ; s = \sum_{\lambda=1}^{m} s_\lambda \omega_\lambda$$

Let $A = \left(a_{ij}\right)_{\substack{i=1...r \\ j=1...r}}$ a matrix of elements in $F_{2^m}$

Let the system $Av^t = s$

For $i = 1...r$ it was

$$s_i = \sum_{j=1}^{n} a_{ij} v_j = \sum_{j=1}^{n} a_{ij} \left( \sum_{\lambda=1}^{m} v_j^\lambda \omega_\lambda \right) = \sum_{\lambda=1}^{m} \left( \sum_{j=1}^{n} a_{ij} v_j^\lambda \right) \omega_\lambda = \sum_{\lambda=1}^{m} s_i^\lambda \omega_\lambda$$

therefore $s_i^\lambda = \sum_{j=1}^{n} a_{ij} v_j^\lambda$ we conclude that

$\forall i = 1, ..., n \; v_i \in F_{2^m}$ we put $v_i = \sum_{\lambda=1}^{m} v_i^\lambda \omega_\lambda$

$\forall j = 1, ..., 2r - 1, \; s_j = \sum_{\lambda=1}^{m} s_j^\lambda \omega_\lambda$ therefore

$$A. \begin{bmatrix} v_1^1 & ... & v_1^m \\ . & ... & . \\ v_n^1 & ... & v_n^m \end{bmatrix} = \begin{bmatrix} s_0^1 & ... & s_0^m \\ . & ... & . \\ s_{r-1}^1 & ... & s_{r-1}^m \end{bmatrix}.$$

## VII- SOLVING SYSTEMS OF VANDERMONDE MATRIX $F_2^n$

Let the system $A. \begin{bmatrix} v_1 \\ ... \\ v_n \end{bmatrix} = \begin{bmatrix} S_0 \\ ... \\ S_{2^m-2} \end{bmatrix}$ ; if $A$ is of the form

$$A = \begin{pmatrix} 1 & 1 & ... & 1 \\ \alpha_1 & \alpha_2 & ... & \alpha_n \\ .. & .... & ... & ... \\ \alpha_1^{2^m-2} & \alpha_2^{2^m-2} & ... & \alpha_n^{2^m-2} \end{pmatrix}$$

Suppose that the indexes $i_1, ..., i_k$ of vector $\begin{bmatrix} v_1 \\ ... \\ ... \\ v_n \end{bmatrix}$ are not zero that is to say $v_{i_1} = ... = v_{i_k} = 1$; it follows

$$S_0 = 1 + ... + 1$$
$$S_1 = \alpha_{i_1} + ... + \alpha_{i_k}$$
$$S_2 = \alpha_{i_1}^2 + ... + \alpha_{i_k}^2$$
$$...$$
$$S_{2^m-2} = \alpha_{i_1}^{2^m-2} + ... + \alpha_{i_k}^{2^m-2}$$

We pose $\alpha_{i_1} = x_1, ..., \alpha_{i_k} = x_k$ therefore

$$S_i = \sum_{j=1}^{k} x_j^i \ , i = 0,1,...,2^m - 2$$

Let $\sigma(x) = \prod_{j=1}^{k}(x - x_j) = \sum_{j=0}^{k} \sigma_j x^{k-j}$

$$v_i = 1 \Leftrightarrow \sigma(\alpha_i) = 0$$

Just find the polynomial $\sigma(x)$ and exhaustive search method known as one dog finds its roots and can be detected by following the indexes $i_1, ..., i_k$.

According to Newton's identity we have:

$$\begin{bmatrix} S_{2^m-2} & S_{2^m-3} & \cdots & S_0 \\ S_0 & S_{2^m-2} & \cdots & S_1 \\ \cdots & \cdots & \cdots & \cdots \\ S_{2^m-3} & S_{2^m-4} & \cdots & S_{2^m-2} \end{bmatrix} \begin{bmatrix} 0 \\ \cdots \\ 0 \\ 1 \\ \sigma_1 \\ \cdots \\ \sigma_k \end{bmatrix} = 0$$

; so we have to solve this system, first study the uniqueness of solution.

***Lemma***

$$rg \begin{bmatrix} S_{2^m-2} & S_{2^m-3} & \cdots & S_0 \\ S_0 & S_{2^m-2} & \cdots & S_1 \\ \cdots & \cdots & \cdots & \cdots \\ S_{2^m-3} & S_{2^m-4} & \cdots & S_{2^m-2} \end{bmatrix} = k$$

***Proof***

$$S_i = \sum_{j=1}^{k} x_j^i = \sum_{j=1}^{n} v_j \alpha_j^i \ , \forall i \in N$$

$$\begin{bmatrix} S_i \\ S_{i+1} \\ \cdots \\ S_{i+2^m-1} \end{bmatrix} = \begin{bmatrix} 1 & 1 & \cdots & 1 \\ \alpha_1 & \alpha_2 & \cdots & \alpha_n \\ . & . & \cdots & . \\ \alpha_1^{2^m-1} & \alpha_2^{2^m-1} & \cdots & \alpha_n^{2^m-1} \end{bmatrix} \begin{bmatrix} v_1\alpha_1^i \\ v_2\alpha_2^i \\ \cdots \\ v_n\alpha_n^i \end{bmatrix}$$

$$C_S = \begin{bmatrix} S_{2^m-2} & S_{2^m-3} & \cdots & S_0 \\ S_0 & S_{2^m-2} & \cdots & S_1 \\ \cdots & \cdots & \cdots & \cdots \\ S_{2^m-3} & S_{2^m-4} & \cdots & S_{2^m-2} \end{bmatrix} = \begin{bmatrix} 1 & 1 & \cdots & 1 \\ \alpha_1 & \alpha_2 & \cdots & \alpha_n \\ & & \cdots & . \\ . & . & . & . \\ \alpha_1^{2^m-1} & \alpha_2^{2^m-1} & \cdots & \alpha_n^{2^m-1} \end{bmatrix} \begin{bmatrix} v_1\alpha_1^{2^m-1} & \cdots & v_1\alpha_1 & v_1 \\ v_2\alpha_2^{2^m-1} & \cdots & v_2\alpha_2 & v_2 \\ \cdots & \cdots & \cdots & \cdots \\ v_n\alpha_n^{2^m-1} & \cdots & v_n\alpha_n & v_n \end{bmatrix}$$

$$= F \begin{bmatrix} v_1 & 0 & . & 0 \\ 0 & v_2 & .. & . \\ \cdots & \cdots & \cdots & 0 \\ 0 & . & ...0 & v_n \end{bmatrix} F'$$

with $F$ and $F'$ two matrix Vandermonde invertible therefore

$$rg(C_S) = card\{j \in \{1,...,n\} / v_j = 1\} = k$$

***Lemma***

For all $i = 0,1,...,2^m - 2$ , $S_i = \sum_{j=1}^{n} v_j \alpha_j^i$ and

$$\sigma(x) = \sum_{j=0}^{k} \sigma_j x^{k-j}$$

The solution $\begin{bmatrix} 0 \\ \cdots \\ 0 \\ 1 \\ \sigma_1 \\ \cdots \\ \sigma_k \end{bmatrix}$ of system

$$\begin{bmatrix} S_{2^m-2} & S_{2^m-3} & \cdots & S_0 \\ S_0 & S_{2^m-2} & \cdots & S_1 \\ \cdots & \cdots & \cdots & \cdots \\ S_{2^m-3} & S_{2^m-4} & \cdots & S_{2^m-2} \end{bmatrix} \begin{bmatrix} 0 \\ \cdots \\ 0 \\ 1 \\ \sigma_1 \\ \cdots \\ \sigma_k \end{bmatrix} = 0$$

of unknown $\begin{bmatrix} 0 \\ \cdots \\ 0 \\ 1 \\ \sigma_1 \\ \cdots \\ \sigma_k \end{bmatrix}$ is unique.

***Proof***

Repeat the same proof as above we obtain the lemma

$$\begin{bmatrix} S_k & S_{k-1} & \cdots & S_0 \\ S_{k+1} & S_k & \cdots & S_1 \\ \cdots & \cdots & \cdots & \cdots \\ S_{2k-1} & S_{2k-2} & \cdots & S_{k-1} \end{bmatrix} = \begin{bmatrix} 1 & 1 & \cdots & 1 \\ \alpha_1 & \alpha_2 & \cdots & \alpha_n \\ & & \cdots & . \\ . & . & . & . \\ \alpha_1^{k-1} & \alpha_2^{k-1} & \cdots & \alpha_n^{k-1} \end{bmatrix} = \begin{bmatrix} v_1 & 0 & . & 0 \\ 0 & v_2 & .. & . \\ \cdots & \cdots & \cdots & 0 \\ 0 & . & ...0 & v_n \end{bmatrix} \begin{bmatrix} \alpha_1^{k-1} & \cdots & \alpha_1 & 1 \\ \alpha_2^{k-1} & \cdots & \alpha_2 & 1 \\ \cdots & \cdots & \cdots & \cdots \\ \alpha_n^{k-1} & \cdots & \alpha_n & 1 \end{bmatrix}$$

So $rg \begin{bmatrix} S_k & S_{k-1} & \cdots & S_0 \\ S_{k+1} & S_k & \cdots & S_1 \\ \cdots & \cdots & \cdots & \cdots \\ S_{2k-1} & S_{2k-2} & \cdots & S_{k-1} \end{bmatrix} = k$ then $\begin{bmatrix} S_{k-1} & S_{k-2} & \cdots & S_0 \\ S_k & S_{k-1} & \cdots & S_1 \\ \cdots & \cdots & \cdots & \cdots \\ S_{2k-2} & S_{2k-3} & \cdots & S_{k-1} \end{bmatrix}$

is invertible.

$$C_S \cdot \begin{bmatrix} 0 \\ \cdots \\ 0 \\ 1 \\ \sigma_1 \\ \cdots \\ \sigma_k \end{bmatrix} = 0 \Rightarrow \begin{bmatrix} S_k & S_{k-1} & \cdots & S_0 \\ S_{k+1} & S_k & \cdots & S_1 \\ \cdots & \cdots & \cdots & \cdots \\ S_{2k-1} & S_{2k-2} & \cdots & S_{k-1} \end{bmatrix} \begin{bmatrix} 1 \\ \sigma_1 \\ . \\ . \\ . \\ \sigma_k \end{bmatrix} = 0 \Rightarrow \begin{bmatrix} S_{k-1} & S_{k-1} & \cdots & S_0 \\ S_k & S_k & \cdots & S_1 \\ \cdots & \cdots & \cdots & \cdots \\ S_{2k-2} & S_{2k-3} & \cdots & S_{k-1} \end{bmatrix} \begin{bmatrix} \sigma_1 \\ \sigma_2 \\ . \\ . \\ . \\ \sigma_k \end{bmatrix} = \begin{bmatrix} S_k \\ S_{k+1} \\ . \\ . \\ . \\ S_{2k-1} \end{bmatrix}$$

Since $\begin{bmatrix} S_{k-1} & S_{k-1} & \cdots & S_0 \\ S_k & S_k & \cdots & S_1 \\ \cdots & \cdots & \cdots & \cdots \\ S_{2k-2} & S_{2k-3} & \cdots & S_{k-1} \end{bmatrix}$ is invertible then $\begin{bmatrix} \sigma_1 \\ \sigma_2 \\ . \\ . \\ \sigma_k \end{bmatrix}$ is unique.

***Proposition***

For all $i = 0,1,\ldots,2^m - 2$, $S_i = \sum_{j=1}^{n} v_j \alpha_j^i$ and

$\sigma(x) = \sum_{j=0}^{k} \sigma_j x^{k-j}$ and

$Q(x) = S_{2^m-2} + S_{2^m-3}x + \ldots + S_1 x^{2^m-3} + S_0 x^{2^m-2}$

$$\begin{bmatrix} S_{2^m-2} & S_{2^m-3} & \cdots & S_0 \\ S_0 & S_{2^m-2} & \cdots & S_1 \\ \cdots & \cdots & \cdots & \cdots \\ S_{2^m-3} & S_{2^m-4} & \cdots & S_{2^m-2} \end{bmatrix} \begin{bmatrix} 0 \\ \cdots \\ 0 \\ 1 \\ \sigma_1 \\ \cdots \\ \sigma_k \end{bmatrix} = 0 \Leftrightarrow Q(\alpha^i)\sigma(\alpha^i) = 0$$

$; \forall i = 1,\ldots,2^m - 2$

***Proof***

By lemma 4 we have

$$\begin{bmatrix} S_{2^m-2} & S_{2^m-3} & \cdots & S_0 \\ S_0 & S_{2^m-2} & \cdots & S_1 \\ \cdots & \cdots & \cdots & \cdots \\ S_{2^m-3} & S_{2^m-4} & \cdots & S_{2^m-2} \end{bmatrix} = \begin{bmatrix} 1 & 1 & \cdots & 1 \\ 1 & \alpha & \cdots & \alpha^{2^m-2} \\ . & . & \cdots & . \\ . & . & . & . \\ . & . & . & . \\ 1 & \alpha^{(2^m-2)} & \cdots & \alpha^{(2^m-2)(2^m-2)} \end{bmatrix} \begin{bmatrix} Q(1) & & & \\ & Q(\alpha) & & \\ & & \ddots & \\ & & & Q(\alpha^{2^m-2}) \end{bmatrix} \begin{bmatrix} 1 & 1 & \cdots & 1 \\ 1 & \alpha^{2^m-2} & \cdots & \alpha \\ . & . & \cdots & . \\ . & . & . & . \\ . & . & . & . \\ 1 & \alpha^{(2^m-2)(2^m-2)} & \cdots & \alpha^{(2^m-2)} \end{bmatrix}$$

With $Q(x) = S_{2^m-2} + S_{2^m-3}x + \ldots + S_1 x^{2^m-3} + S_0 x^{2^m-2}$

We will

$$\begin{bmatrix} S_{2^m-2} & S_{2^m-3} & \cdots & S_0 \\ S_0 & S_{2^m-2} & \cdots & S_1 \\ \cdots & \cdots & \cdots & \cdots \\ S_{2^m-3} & S_{2^m-4} & \cdots & S_{2^m-2} \end{bmatrix} \begin{bmatrix} 0 \\ \cdots \\ 0 \\ 1 \\ \sigma_1 \\ \cdots \\ \sigma_k \end{bmatrix} = P \begin{bmatrix} Q(1) & & & \\ & Q(\alpha) & & \\ & & \ddots & \\ & & & Q(\alpha^{2^m-2}) \end{bmatrix} P^{-1} \begin{bmatrix} 0 \\ \cdots \\ 0 \\ 1 \\ \sigma_1 \\ \cdots \\ \sigma_k \end{bmatrix}$$ therefore

$$C_S \cdot \begin{bmatrix} 0 \\ \cdots \\ 0 \\ 1 \\ \sigma_1 \\ \cdots \\ \sigma_k \end{bmatrix} = 0 \Leftrightarrow \begin{bmatrix} Q(1) & & & \\ & Q(\alpha) & & \\ & & \ddots & \\ & & & Q(\alpha^{2^m-2}) \end{bmatrix} \begin{bmatrix} 1 & 1 & \cdots & 1 \\ 1 & \alpha^{2^m-2} & \cdots & \alpha \\ . & . & \cdots & . \\ . & . & . & . \\ . & . & . & . \\ 1 & \alpha^{(2^m-2)(2^m-2)} & \cdots & \alpha^{(2^m-2)} \end{bmatrix} \begin{bmatrix} 0 \\ \cdots \\ 0 \\ 1 \\ \sigma_1 \\ \cdots \\ \sigma_k \end{bmatrix} = 0$$

$$\Leftrightarrow \begin{bmatrix} Q(1) & & & \\ & Q(\alpha) & & \\ & & \ddots & \\ & & & Q(\alpha^{2^m-2}) \end{bmatrix} \begin{bmatrix} \sigma_k + \ldots + \sigma_1 + 1 \\ \sigma_k \alpha + \sigma_{k-1}\alpha^2 + \ldots + \sigma_1 \alpha^k + 1 \\ \cdots \\ \cdots \\ \cdots \\ \sigma_k \alpha^{2^m-2} + \sigma_{k-1}\alpha^{2(2^m-2)} + \ldots + \sigma_1 \alpha^{k(2^m-2)} + 1 \end{bmatrix} = 0$$

$$\Leftrightarrow \begin{bmatrix} Q(1)\sigma(1) \\ Q(\alpha)\sigma(\alpha)\alpha \\ \cdots \\ Q(\alpha^{2^m-2})\sigma(\alpha^{2^m-2})\alpha^{2^m-2} \end{bmatrix} = 0 \Leftrightarrow Q(\alpha^i)\sigma(\alpha) = 0$$

for all $i = 0,1,\ldots,2^m - 2$

***Proposition***

We put for $i = 0,1,\ldots,2^m - 2$, $S_i = \sum_{j=1}^{n} v_j \alpha_j^i$ and

$\sigma(x) = \sum_{j=0}^{k} \sigma_j x^{k-j}$ and

$Q(x) = S_{2^m-2} + S_{2^m-3}x + \ldots + S_1 x^{2^m-3} + S_0 x^{2^m-2}$

$\{i, \sigma(\alpha^i) = 0\} = \{i, Q(\alpha^i) \neq 0\} = k$ and

$Q(\alpha^i) \neq 0 \Leftrightarrow \sigma(\alpha^i) = 0$

***Proof***

By the previous proposal it was

$Q(\alpha^i)\sigma(\alpha^i) = 0 \;; \forall i = 1,\ldots,2^m - 2$ therefore if

$Q(\alpha^i) \neq 0$ we will $\sigma(\alpha^i) = 0$ then

$\{i, Q(\alpha^i) \neq 0\} \subset \{i, \sigma(\alpha^i) = 0\}$ gold

$card\{i, Q(\alpha^i) \neq 0\} = rgC_S = k$ and

$card\{i, \sigma(\alpha^i) = 0\} = k$

## VIII – Our decoding algorithm

The syndrome vector all received word not exceeding the correction capacity is calculated by simple matrix product control word received by the. We still have to find a method to calculate the extended syndromes. We must convert each element of $F_{2^m}$ a column vector $m$ component of $F_2$ taken with respect to a natural base $\{1, \alpha, \ldots, \alpha^{m-1}\}$.

Algorithm

Input : $(S_j^{\lambda})_{\substack{j=0,1,\ldots 2^m-2 \\ \lambda=1,\ldots m}}$

Output : $(e_1, \ldots, e_n)$

for $\lambda = 1,\ldots,m$ $Q^{\lambda}(x) = S_{2^m-2}^{\lambda} + S_{2^m-3}^{\lambda}x + \ldots + S_1^{\lambda}x^{2^m-3} + S_0^{\lambda}x^{2^m-2}$

for $i = 1,\ldots,n$ $Q^{\lambda}(\alpha_i) \neq 0$ these $e_i^{\lambda} = 1$ else $e_i^{\lambda} = 0$

for $i = 1,\ldots,n$ $F_2^m \to F_{2^m}$ $(e_i^1, \ldots, e_i^m) \to E_i$ and if

$E_i = 0$ then $e_i = 0$ else $e_i = 1$

## IX- Conclusion

Our approach overcomes the vulnerability of cryptosystems MC Eliece, incured as information leakage, caused by the fact that the number of iterations in the Euclidean algorithm is influenced by the number of error bits that this cryptosystem must hide. This approach requires to find an effective method for calculate the syndromes of extended classical irreducible Goppa codes.

## X- bibliographie